\documentclass[aps,prb,twocolumn,superscriptaddress,nofootinbib]{revtex4-2}

\usepackage[T1]{fontenc}
\usepackage[utf8]{inputenc}
\usepackage{amsmath,amssymb,bm}
\usepackage{mathrsfs}
\usepackage{graphicx}
\usepackage{booktabs}
\usepackage{tabularx}
\usepackage{array}
\usepackage{xcolor}
\usepackage{hyperref}
\hypersetup{colorlinks=true,linkcolor=blue,citecolor=blue,urlcolor=blue}

\newif\ifshowchanges
\showchangesfalse
\newcommand{\rev}[1]{\ifshowchanges\textcolor{red}{#1}\else#1\fi}

\showchangesfalse

\begin{document}

\title{Generalized Space Groups from Internal Configuration Spaces}
\author{Zeying Zhang} 
\email{zzy@mail.buct.edu.cn}
\affiliation{College of Mathematics and Physics, Beijing University of Chemical
	Technology, Beijing 100029, China}

\author{Zhenye Li}
\email{lizhenye@pku.edu.cn}
\affiliation{College of Mathematics and Physics, Beijing University of Chemical
	Technology, Beijing 100029, China}
\author{Zhi-Ming Yu}
\affiliation{Centre for Quantum Physics, Key Laboratory of Advanced Optoelectronic Quantum Architecture and Measurement (MOE),
	Beijing Institute of Technology, Beijing 100081, China}
\author{Gui-Bin Liu}
\affiliation{Centre for Quantum Physics, Key Laboratory of Advanced Optoelectronic Quantum Architecture and Measurement (MOE),
	Beijing Institute of Technology, Beijing 100081, China}
\author{Yugui Yao}
\email{ygyao@bit.edu.cn}
\affiliation{Centre for Quantum Physics, Key Laboratory of Advanced Optoelectronic Quantum Architecture and Measurement (MOE),
	Beijing Institute of Technology, Beijing 100081, China}
	
\begin{abstract}
We develop a unified construction of generalized space groups for crystals with unconventional internal degrees of freedom. Starting from the full group $G_P$ of allowed internal transformations and the stabilizer $P$ of a reference object, we determine the pointwise and setwise symmetries, $J$ and $K$, of the allowed configuration set. Goursat's lemma then couples the internal quotient $K/J$ to a spatial quotient. The framework includes ordinary, magnetic, spin, and color space groups as special cases. As an example, we consider a dodecahedral object with $P=I\simeq A_5$, for which we obtain the nontrivial pair $T\triangleleft O$ with $O/T\simeq\mathbb Z_2$. The resulting generalized space group hosts a point node with topological charge $|C|=12$.
\end{abstract}

\maketitle

\textit{Introduction}---Space groups (SGs) provide the fundamental symmetry classification of crystalline solids. They describe how spatial rotations, reflections, and translations map a crystal onto itself, and form the basis for analyzing crystal structures, lattice vibrations, electronic bands, and symmetry-protected topology \cite{bradley_mathematical_2009, lax_symmetry_2001, dresselhaus_group_2008, bansil_colloquium_2016, chiu_classification_2016, cano_band_2021, yu_encyclopedia_2022}. In an ordinary space group, each atomic site is characterized only by its position and chemical identity, while every symmetry operation acts solely on real space. A crystalline site may, however, carry additional internal attributes. When the local object is a magnetic moment that changes sign under time reversal, combining spatial operations with time reversal leads to magnetic space groups (MSGs) \cite{opechowski_magnetic_1965,bradley_magnetic_1968, litvin_magnetic_2016}. If the full orientation of a spin arrow is retained and spin rotations can be combined independently with spatial operations, the resulting symmetry is described by spin space groups \cite{w_f_brinkman_theory_1966,litvin_spin_1974,litvin_spin_1977,chen_enumeration_2024,jiang_enumeration_2024,xiao_spin_2024}. Moreover, a set of discrete labels may be attached to crystallographically equivalent sites and permuted by symmetry operations, giving rise to color groups (CGs) \cite{van_der_waerden_farbgruppen_1961, av_shubnikov_colored_1964, schwarzenberger_colour_1984, lifshitz_theory_1997}. These extensions share a common idea: a spatial operation need not preserve an internal configuration by itself, provided that it is accompanied by an appropriate transformation of the internal degree of freedom.

This observation raises a more general question: what symmetry groups arise when a local object carries an internal attribute other than time reversal, a vector, or a prescribed set of colors? Relevant settings include orientational order in molecular crystals \cite{heiney_orientational_1991,bostrom_recipes_2018}, local resonators in sonic, optomechanical, and mechanical metamaterials \cite{liu_locally_2000,eichenfield_optomechanical_2009,nash_topological_2015}, internal configurations of skyrmions \cite{lin_internal_2014,zhang_skyrmion_2017}, orbital and multipole order \cite{murakami_resonant_1998,watanabe_group-theoretical_2018}, and composite objects in cluster crystals \cite{lenz_microscopically_2012,stiakakis_self_2021}. Metal--organic frameworks offer a particularly rich realization, combining designed networks of metal-based and organic building units \cite{yaghi_reticular_2003,li_topological_2014,glasby_topological_2024} with orientational, spin-state, ligand, and guest ordering \cite{mariette_symmetry_2020}. The corresponding internal operation group need not be a subgroup of $O(3)$; it may be any group $G_P$ acting on a configuration space $X$. Existing classifications for particular internal degrees of freedom do not address this general setting, while a purely algebraic enumeration includes subgroups that need not be realized by any allowed configuration set.

In this work, we develop a general construction of generalized space groups (GSGs) starting from only the full group $G_P$ of allowed internal transformations and the symmetry $P$ of a reference internal object. The algebraic coupling between a spatial group $G_S$ and an internal group is described using Goursat's lemma. For full projections onto both factors, the coupling is determined by normal subgroups and an isomorphism between the corresponding quotient groups. The remaining physical problem is then to determine which internal quotient can actually be realized by the allowed configurations.

\textit{Construction principle}---Let $G_S$ be an ordinary space group acting on the positions of sites,
and let $G_P$ be a group of allowed transformations acting on an
internal configuration space $X$. If $\chi(\bm r)$ denotes the internal
configuration assigned to the site at $\bm r$, the spatial and internal
transformations act as
$
[(g,p)\chi](\bm r)=p\,\chi(g^{-1}\bm r);  (g,p)\in G_S\times G_P.
$
 The symmetry group $\mathcal G$ of the decorated crystal is therefore a subgroup of 
$G_S\times G_P.$
We restrict attention to structures whose spatial projection is the full parent space group, i.e $\{g\mid (g,p)\in\mathcal G\}=G_S$.
The internal projection is defined by
$
S_P=\{p\mid (g,p)\in\mathcal G\}\le G_P.
$
Consequently, $\mathcal G$ is a subgroup of $G_S\times S_P$ with surjective projections onto both factors. Goursat's lemma characterizes all subgroups of
$G_S\times S_P$ with surjective projections. That is, let
$
N_S\triangleleft G_S
$
and
$
N_P\triangleleft S_P,
$
together with an isomorphism
$
\varphi:G_S/N_S\longrightarrow S_P/N_P
$ \cite{hall_theorem_1963}.
Explicitly,
\begin{equation}
\mathcal G=
\{(g,p)\in G_S\times S_P\mid
\varphi(gN_S)=pN_P\}.
\label{eq:goursat-subgroup}
\end{equation}
Since $S_P\le G_P$, the resulting group $\mathcal G$ is also a subgroup of $G_S\times G_P$, with internal projection $S_P$. Thus Goursat's lemma classifies all possible GSGs once $S_P$ has been specified. The remaining problem is to determine how the allowed internal configurations select the admissible $S_P$.

\begin{figure}
	\centering
	\includegraphics[width=\columnwidth]{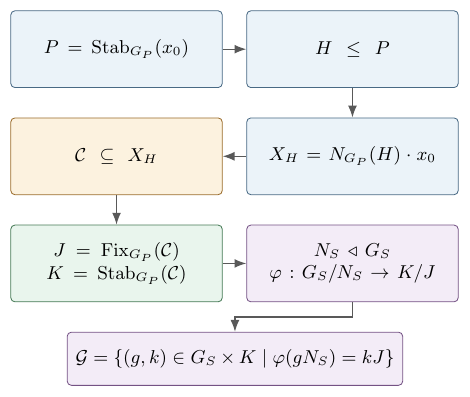}
	\caption{Construction of a generalized space group. The preset subgroup $H$ generates the candidate space $X_H$, while the physically selected subset $\mathcal C$ determines $J\triangleleft K$. Goursat's lemma then couples the internal quotient $K/J$ to a spatial quotient $G_S/N_S$.}
	\label{fig:gsg-construction}
\end{figure}

 In principle, one could enumerate all subgroups $S_P\le G_P$ and apply the construction to each of them. Such an enumeration, however, is not yet a physical classification, because only those $S_P$ compatible with the allowed internal configurations are physically relevant.
For example,  in SSGs, 
$
G_P=SO(3)\times\mathbb Z_2^T.
$
For a collinear magnetic configuration along $\hat n$, the maximal internal symmetry is
$
S_P=C_{\infty v}=SO(2)\rtimes\mathbb Z_2^{C_2T},
$
rather than any finite axial subgroup $C_n$.
The admissible internal projection should therefore be determined from the full symmetry of the allowed configuration set, rather than chosen as an arbitrary subgroup of $G_P$. More generally, an algebraically available subgroup may either retain only a nonmaximal part of the configuration symmetry or map an allowed configuration outside the physically relevant set.

We next determine the internal projection $S_P$ from the allowed internal configurations. Let $G_P$ act on a configuration space $X$. The stabilizer of each $x\in X$ is
$
\operatorname{Stab}_{G_P}(x)
=\{p\in G_P\mid px=x\}.
$
For a finite configuration set
$\mathcal C=\{x_1,\ldots,x_m\}\subset X$, its setwise and pointwise stabilizers are
\begin{equation}
\begin{aligned}
K&=\operatorname{Stab}_{G_P}(\mathcal C)
=\{p\in G_P\mid p\mathcal C=\mathcal C\},\\
J&=\bigcap_{i=1}^m\operatorname{Stab}_{G_P}(x_i),
\end{aligned}
\label{eq:JK-stabilizers}
\end{equation}
respectively. The group $K$ preserves $\mathcal C$ as a whole and may permute its elements, whereas $J$ fixes every configuration individually. Hence, $J\triangleleft K$, and the quotient group $K/J$ acts faithfully on $\mathcal C$ by permuting its elements. The maximal internal projection determined by $\mathcal C$ is therefore
\begin{equation}
S_P=K,\qquad N_P=J.
\label{eq:maximal-internal-data}
\end{equation}
Here, ``maximal'' means retaining the full symmetry of $\mathcal C$.
Although one may choose a proper normal subgroup $N_P\subsetneq J$,
this merely discards internal operations that fix every configuration
and yields a nonmaximal subgroup. Different physical symmetries are
instead obtained by varying $\mathcal C$ and recomputing $J$ and $K$.

\begin{table*}
		\caption{Examples of GSGs. The notation follows that introduced in the main text. Here, $R\leq\operatorname{Sym}(n)$ is a regular permutation
			group on the $n$ colors. \rev{For DGSGs, $SO(3)/I$ is the
			configuration space of all distinct dodecahedral orientations,
			while $D_\infty/D_n$ is the candidate space of distinct orientations
			sharing a preset $C_n$ axis; in each case.} In the SSG and DGSG rows with
			$J=1,C_n,C_s$, the listed configuration sets exclude the cases covered
			by the more specialized rows.}
		\label{tab:gsg-constructions}
	\scriptsize
	\setlength{\tabcolsep}{3pt}
	\resizebox{\textwidth}{!}{%
		\begin{tabular}{lllllllll}
			\toprule
			Group & $G_P$ & $P$ & $H$ & $\mathcal C$
			& $J$ & $K$ & $K/J$ &  System \\
			\midrule
			SG
			& $1$
			& $1$
			& $1$
			& $\{x_0\}$
			& $1$
			& $1$
			& $1$
			& ordinary space group \\
			\midrule
			MSG
			& $\mathbb Z_2^T$
			& $1$
			& $1$
			& $\{\mathrm{black}\}$
			& $1$
			& $1$
			& $1$
			& type-I MSG \\
			
			&
			&
			& $1$
			& $\{\mathrm{black},\mathrm{white}\}$
			& $1$
			& $\mathbb Z_2^T$
			& $\mathbb Z_2$
			& type-III or type-IV MSG \\
			
			&
			& $\mathbb Z_2^T$
			& $\mathbb Z_2^T$
			& $\{\mathrm{gray}\}$
			& $\mathbb Z_2^T$
			& $\mathbb Z_2^T$
			& $1$
			& type-II MSG \\
			
			\midrule
			CG
			& $R\leq\operatorname{Sym}(n)$
			& $1$
			& $1$
			&  set of $n-$colors$$
			& $1$
			& $K\leq R$
			& $K$
			& CG with $n$ colors\\
			\midrule
			SSG
			& $O(3)$
			& $C_{\infty v}$
			& $C_n$
			& $\{\hat{ n}\}$
			& $C_{\infty v}$
			& $C_{\infty v}$
			& $1$
			& collinear FM \\
			
			&
			&
			&$C_n$
			& $\{\hat{ n},-\hat{ n}\}$
			& $C_{\infty v}$
			& $D_{\infty h}$
			& $\mathbb Z_2$
			& collinear AFM or AM \\
			&
			&
			&
			$C_s$
			& $\{x_1,\ldots,x_m\}\subset S^1$
			& $C_s$
			& $\operatorname{Stab}_{O(3)}(\mathcal C)$
			& $K/C_s$
			& finite coplanar order \\
			
			&
			&
			& $1$
			& $\{x_1,\ldots,x_m\}\subset S^2$
			& $1$
			& $\operatorname{Stab}_{O(3)}(\mathcal C)$
			& $K$
			& finite noncoplanar order \\					
			\midrule

			DGSG
& $SO(3)$
& $I$
&$I$
& $\{x_0\}$
& $I$
& $I$
& $1$
& single configuration \\

			%
			%
			%

			
			&
			&
			&
			$D_3$
			& $\{x_1,x_2\}$
			& $D_3$
			& $D_6$
			& $\mathbb{Z}_2$
			& binary dodecahedral \\
			
			&
			&
			&
			$D_5$
			& $\{x_1,x_2\}$
			& $D_5$
			& $D_{10}$
			& $\mathbb{Z}_2$
			& binary dodecahedral \\
			
			&
			&
			&
			$D_2, T$
			& $\{x_1,x_2\}$
			& $T$
			& $O$
			& $\mathbb{Z}_2$
			& binary dodecahedral \\
			
		&
&
&
$C_n$
& $\mathcal C\subseteq D_\infty/D_n$
& $C_n$
& $\operatorname{Stab}_{SO(3)}(\mathcal C)$
& $K/C_{n}$
& finite axial dodecahedral order \\
			
& 
& 
& 1
& $\mathcal C\subseteq SO(3)/I$
& 1
& $\operatorname{Stab}_{SO(3)}(\mathcal C)$
&$K$
& general orientations \\

			\midrule
			
			& $G_P$
			& $\operatorname{Stab}_{G_P}(x_0)$
			& $H\le P$
			& $\mathcal C\subseteq X_H$
			& $\operatorname{Fix}_{G_P}(\mathcal C)$
			& $\operatorname{Stab}_{G_P}(\mathcal C)$
			& $K/J$
			& general construction \\
			\bottomrule
		\end{tabular}%
	}
\end{table*}

It remains to determine the admissible configuration sets $\mathcal C$, whose elements and common symmetries are not known a priori. To construct them systematically, we choose a reference configuration $x_0$ with stabilizer
$
P=\operatorname{Stab}_{G_P}(x_0),
$
and consider each subgroup $H\le P$ as a candidate common symmetry.
For each $H$, let
$
N_{G_P}(H)=\{p\in G_P\mid pHp^{-1}=H\},$ and $
N_P(H)=\{p\in P\mid pHp^{-1}=H\}=N_{G_P}(H)\cap P
$
be the normalizers of $H$ in $G_P$ and $P$, respectively. The first generates configurations sharing the preset symmetry $H$, while the second leaves the reference configuration $x_0$ unchanged. Hence, the distinct candidate configurations constitute the orbit
\begin{equation}
	X_H=N_{G_P}(H)\cdot x_0.
	\label{eq:candidate-space}
\end{equation}
Since the stabilizer of $x_0$ in $N_{G_P}(H)$ is $N_P(H)$, these
configurations are in one-to-one correspondence with the left cosets:
\begin{equation*}
	\begin{aligned}
N_{G_P}(H)/N_P(H)&\longrightarrow X_H,\\
uN_P(H)&\longmapsto ux_0.
\end{aligned}
\end{equation*}
Let $u_i\in N_{G_P}(H)$ be representatives of the selected cosets, with $u_1=e$, and define
$$
x_i=u_i x_0.
$$
The stabilizer of $x_i$ is $u_iPu_i^{-1}$. Therefore, for the selected configuration set $\mathcal C=\{x_1,\ldots,x_m\}$, the pointwise stabilizer is
$$
J=\bigcap_i u_iPu_i^{-1}.
$$
Since $u_i\in N_{G_P}(H)$ and $H\le P$, for every $h\in H$,
$
h x_i=h u_i x_0
=u_i(u_i^{-1}hu_i)x_0
=u_i x_0=x_i.
$
Thus $H$ fixes every selected configuration, and hence $H\le J$. The preset subgroup $H$ is therefore only a constraint used to generate candidate configurations, whereas the final pointwise stabilizer $J$ may be larger than $H$.

The normalizer construction yields only the candidate space $X_H$. A physical realization corresponds to a selected subset
$
\mathcal C\subseteq X_H,
$
whose pointwise and setwise stabilizers $J$ and $K$ are given by
Eq.~\eqref{eq:JK-stabilizers}. Thus, $H$ is only a preset constraint and
need not coincide with either $J$ or $K$. Assigning the configurations
in $\mathcal C$ to real-space sites then specifies the coupling between the internal and spatial operations.

For example, consider a collinear spin configuration $x_+$ along
$\hat n$, with
$
G_P=SO(3)\times\mathbb Z_2^T,
P=\operatorname{Stab}_{G_P}(x_+)
=SO(2)\rtimes\mathbb Z_2^{C_{2}T}.
$
Choosing $H=C_2\le P$ with its axis along $\hat n$, one finds
$
N_{G_P}(H)=D_\infty\times\mathbb Z_2^T,
$ and $
N_P(H)=P.
$
Consequently,
$
X_H=\{x_+,x_-\}\simeq N_{G_P}(H)/N_P(H),
$
where $x_-$ is obtained by reversing the spin direction. For
$\mathcal C=\{x_+\}$, one has $J=K=P$, corresponding to a
ferromagnetic (FM) configuration when $x_+$ is assigned to every site. For
$\mathcal C=\{x_+,x_-\}$, one instead obtains
$
J=P,  K=N_{G_P}(H),  K/J\simeq\mathbb Z_2,
$
as shown in Table~\ref{tab:gsg-constructions}. Different real-space assignments of $x_+$ and $x_-$ can then produce
antiferromagnetic (AFM) or altermagnetic (AM) orders
\cite{smejkal_beyond_2022}.

The construction is summarized in Fig.~\ref{fig:gsg-construction}. Starting
from a reference configuration $x_0$ with stabilizer $P$, each subgroup
$H\le P$ generates the candidate space $X_H$ in
Eq.~\eqref{eq:candidate-space}. A physically selected subset
$\mathcal C\subseteq X_H$ determines $J$ and $K$ through
Eq.~\eqref{eq:JK-stabilizers}, and hence the maximal internal data in
Eq.~\eqref{eq:maximal-internal-data}. The resulting GSG is obtained by
matching the quotient $K/J$ to a spatial quotient $G_S/N_S$.
Representative constructions for several choices of $G_P$ and $P$ are
listed in Table~\ref{tab:gsg-constructions}.

\textit{Dodecahedral group  as an example}---We now apply the construction to dodecahedral internal objects and refer
to the resulting groups as dodecahedral generalized space groups
(DGSGs). For a reference dodecahedron $x_0$,
$
G_P=SO(3),
P=I\simeq A_5.
$
Up to conjugacy in $I$, the subgroups of $P$ are \cite{noauthor_gap_2024}:
\[
\{e\},\ C_2,\ C_3,\ C_5,\ D_2,\ D_3,\ D_5,\ T,\ I,
\]
where $T\simeq A_4$, $D_2\simeq C_2\times C_2$, and $D_n$ denotes the
rotational dihedral group of order $2n$. We next determine the candidate
configuration spaces generated by these subgroups and the corresponding
pairs $J\triangleleft K$.

\begin{figure}[t]
	\centering
	\includegraphics[width=\columnwidth]{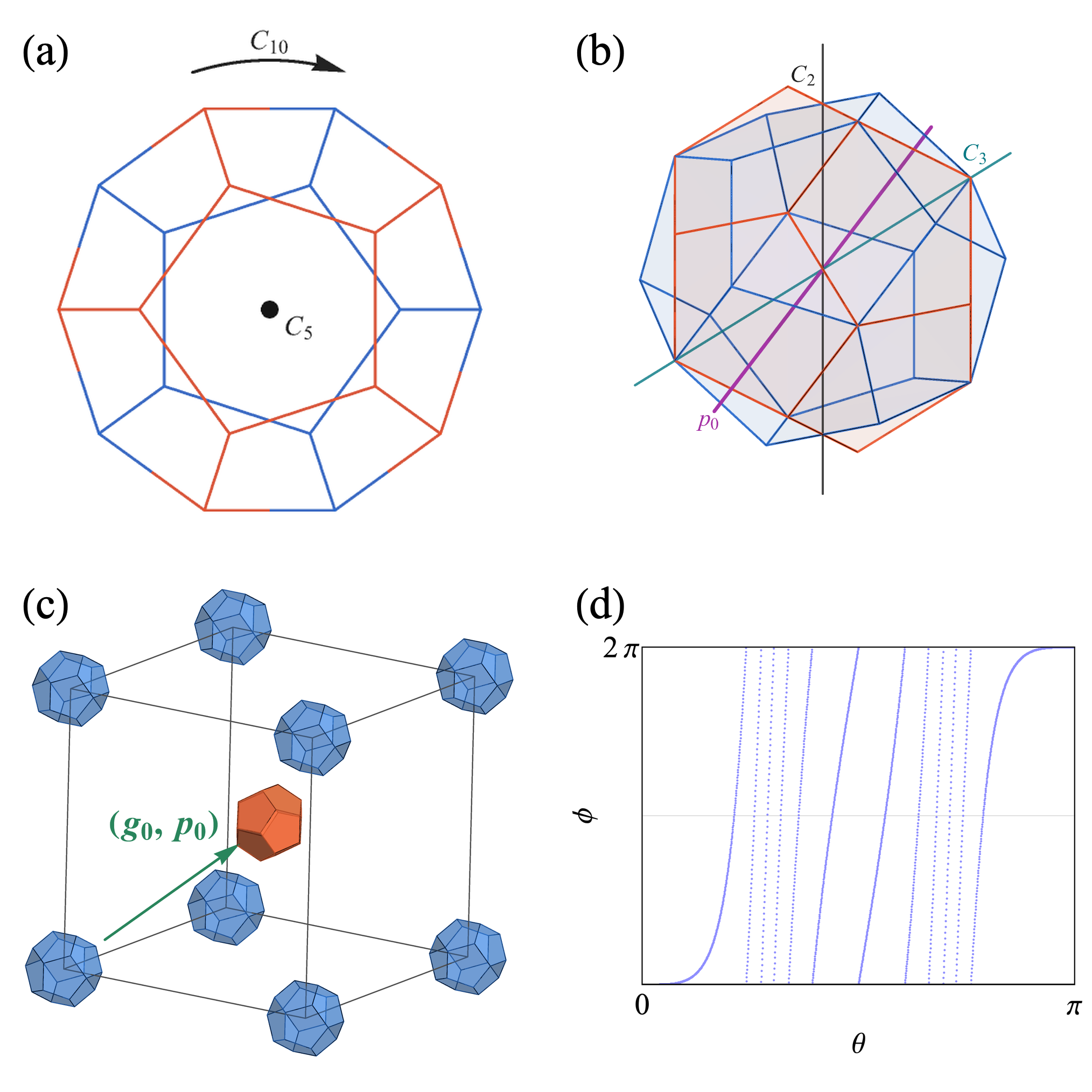}
	\caption{(a) Two dodecahedral orientations sharing a $C_5$ axis and differing by
		a rotation of $\pi/5$. (b) Two configurations associated with
		$T\triangleleft O$. The black and cyan lines mark representative shared
		$C_2$ and $C_3$ axes, and the magenta line marks the axis of
		$p_0=C_{2_{0\bar 1 1}}$. (c) A body-centered realization, with $x_1$
		and $x_2$ on the two sublattices related by $(g_0,p_0)$.
		(d) Wilson-loop spectrum of the six-band $k\cdot p$ model around
		$\Gamma$.}
	\label{fig:2}
\end{figure}

The normalizer quotient separates the subgroups of $I$ into two classes: the trivial and axial subgroups generate continuous candidate spaces,
while the multi-axis subgroups generate discrete configuration sets.
For $H=\{e\}$, the candidate space is the full orientation space
$SO(3)/I$. For an axial subgroup $H=C_n$ ($n=2,3,5$), its normalizer in
$SO(3)$ is $D_\infty\simeq O(2)$, giving a continuous family of
orientations about the common axis. The final pair $J\triangleleft K$
therefore depends on the selected subset $\mathcal C$. For example,
Fig.~\ref{fig:2}(a) shows two configurations sharing a $C_5$ axis and
differing by a rotation of $\pi/5$.

By contrast, the multi-axis subgroups have discrete normalizer quotients.
For $H=D_3$, $D_5$, and $T$, each quotient contains two configurations,
leading respectively to
\[
D_3\triangleleft D_6,\qquad
D_5\triangleleft D_{10},\qquad
T\triangleleft O,
\]
with $K/J\simeq \mathbb{Z}_2$. For $H=D_2$, the two configurations share the
larger pointwise symmetry $T$, and hence give the same pair
$T\triangleleft O$. This illustrates that the preset symmetry $H$ may
be smaller than the resulting pointwise stabilizer $J$.
Fig~\ref{fig:2}(b) shows the two configurations associated with
$T\triangleleft O$ and a representative operation $p_0$ that exchanges
them.

Although the candidate space may be continuous, a periodic ordered
crystal selects only a finite configuration set $\mathcal C$, since
each unit cell contains finitely inequivalent sites. For example,
three equally spaced orientations about a common fivefold axis give
\[
J=C_5,\qquad K=D_{15},\qquad K/J\simeq D_3\simeq S_3.
\]
The perpendicular twofold rotations reverse the common axis and enlarge
the cyclic permutation symmetry to $D_3$. 

As a representative binary case, consider $T\triangleleft O$. The corresponding generalized space group is
$$
\mathcal G
=
\{(g,p)\in G_S\times O\mid
\varphi(gN_S)=pT\}. 
$$
We illustrate the construction using the index-two structure encoded
in MSG $P_I432$ (No.~207.43). The MSG here is used  only
to identify an index-two subgroup. Its antiunitary operation
is not retained in the resulting GSG.
$P_I432$ contains the ordinary space group
$N_S=P432$ (No.~207) as its unitary spatial subgroup. Its spatial
projection is $G_S=I432$ (No.~211).  Adding the body-centering translation
$g_0=\{E|\frac12\!\frac12\!\frac12\!\}$, extends $N_S$ to the spatial projection $I432$. Thus
$$
G_S=N_S\cup g_0N_S,
\quad
G_S/N_S\simeq\mathbb Z_2.
$$
On the internal side, On the $T$ is an index-two normal subgroup of $O$.
Consequently,
\[
O=T\cup p_0T,\quad O/T\simeq\mathbb Z_2,
\]
where $p_0$ is a representative of the nontrivial coset. We choose
$p_0=C_{2_{0\bar 1 1}}$, for which $p_0^2=E$, as shown in Fig.~\ref{fig:2}(b). The quotient isomorphism entering
Goursat's lemma is
$$
\varphi:G_S/N_S\longrightarrow O/T,
\qquad
N_S\mapsto T,
\qquad
g_0N_S\mapsto p_0T.
$$
The resulting GSG is
\begin{equation}
\begin{split}
	\mathcal G
	&=
	\left\{
	(g,p)\in G_S\times O
	\mid
	\varphi(gN_S)=pT
	\right\} \\
	&=
	(N_S\times T)
	\cup
	(g_0,p_0)(N_S\times T).
\end{split}
\label{eq:dodecahedral-gsg}
\end{equation}
\rev{Fig~\ref{fig:2}(c) shows the corresponding body-centered
realization, in which the paired operation $(g_0,p_0)$ exchanges the two
configuration sublattices.}
For each $g\in G_S$, choose the representative
\[
s(g)=
\begin{cases}
	(g,E), & g\in N_S,\\
	(g,p_0), & g\in g_0N_S.
\end{cases}
\]
Their conjugation action on the
internal-only subgroup $\{(e,t)\mid t\in T\}$ defines
$\alpha:G_S\to\operatorname{Aut}(T)$ through $
s(g)(e,t)s(g)^{-1}=(e,\alpha_g(t)).
$
Explicitly,
\[
\alpha_g(t)=
\begin{cases}
	t, & g\in N_S,\\
	p_0tp_0^{-1}, & g\in g_0N_S.
\end{cases}
\]
Therefore,
\[
\mathcal G\simeq T\rtimes_\alpha G_S.
\]
Although $J=T$ plays the role of a spin-only group, it is not an
independent direct-product factor: $N_S$ acts trivially on $T$, whereas
$g_0N_S$ acts by $t\mapsto p_0tp_0^{-1}$, giving
$\mathcal G\simeq T\rtimes_\alpha G_S$.

We calculate the irreducible small representations at $\Gamma$
following the method of Ref.~\cite{zhang_spin_2026}, with the group
computations performed using \textsf{GAP}~\cite{noauthor_gap_2024}.
The generalized little group has 25 irreducible representations: four one-dimensional,
four two-dimensional, eight three-dimensional, one four-dimensional,
four six-dimensional, and four nine-dimensional representations. Remarkably, one of the six-dimensional representations supports
a sixfold chiral node with topological charge $|C|=12$. Using
\textsf{MagneticKP}~\cite{zhang_magnetickp_2023}, its symmetry-allowed
$k\cdot p$ Hamiltonian up to third order in momentum is
$H=H_0\oplus H_0\oplus H_0$, where
\begin{equation*}
	H_0(\bm k)
	=
	\begin{pmatrix}
		c_1k_xk_yk_z
		&
		c_2\left(k_x^2+\omega k_z^2+\omega^2k_y^2\right)
		\\[4pt]
		c_2\left(k_x^2+\omega k_y^2+\omega^2k_z^2\right)
		&
		-c_1k_xk_yk_z
	\end{pmatrix},
		\label{eq:Gamma-kp}
\end{equation*}
with $\omega=e^{2\pi i/3}$ and real parameters $c_1$ and $c_2$.
Each two-band block carries a topological charge of magnitude
$|C_0|=4$ on a closed surface enclosing $\Gamma$. The three
symmetry-related blocks have the same chirality, giving the total charge
$
|C|=3|C_0|=12.
$
\rev{The Wilson-loop spectrum in Fig.~\ref{fig:2}(d) confirms the
total topological charge $|C|=12$.}
To the best of our knowledge, and in comparison with systematic
classifications of band crossings and quasiparticles in ordinary space
groups, magnetic space groups, and spin space groups
\cite{yu_encyclopedia_2022,tang_exhaustive_2021,tang_complete_2022,
liu_systematic_2022,zhang_encyclopedia_2022,yang_symmetry_2024},
a symmetry-enforced point node with $|C|=12$ has not previously been
reported.
This result demonstrates that GSGs can support topological charges
beyond those available in the established crystallographic symmetry
frameworks.

\textit{Conclusion}---we have shown that a broad class of GSGs can be
constructed from only two basic ingredients: an full group $G_P$
acting on an internal configuration space $X$, and the symmetry group $P$ of a reference object $x_0$.
Different physically allowed configuration sets generated from these
data determine different normal pairs $J\triangleleft K$, which can
then be coupled to spatial groups through Goursat's lemma. This
framework contains ordinary SGs, MSGs,
conventional CGs, and SSGs as special cases,
while also producing generalized groups beyond these established classifications. The dodecahedral example with
$P=I\simeq A_5$ explicitly demonstrates that even a familiar finite
object can generate nontrivial generalized space-group structures.

Possible realizations may occur in metal--organic frameworks, molecular and cluster crystals,
orientationally ordered solids, orbital and multipolar systems,
magnetic textures, and photonic, phononic, or mechanical metamaterials.
A complete enumeration of the groups generated by different choices
of $G_P$, $P$, and the physically allowed configuration sets, together
with the identification of their material realizations and band
representations, is a substantial task. The present work provides a
general starting point for such a classification.

	\bibliography{GSG}

\end{document}